\documentclass[a4paper]{jpconf}
\usepackage{graphicx}

\newcommand{\unc}{$^{1}$}
\newcommand{\tunl}{$^{2}$}
\newcommand{\lbnl}{$^{3}$}
\newcommand{\uw}{$^{4}$}
\newcommand{\pnnl}{$^{5}$}
\newcommand{\usc}{$^{6}$}
\newcommand{\ornl}{$^{7}$}
\newcommand{\itep}{$^{8}$}
\newcommand{\usd}{$^{9}$}
\newcommand{\mpi}{$^{10}$}
\newcommand{\jinr}{$^{11}$}
\newcommand{\duke}{$^{12}$}
\newcommand{\sdsmt}{$^{13}$}
\newcommand{\lanl}{$^{14}$}
\newcommand{\ut}{$^{15}$}
\newcommand{\ou}{$^{16}$}
\newcommand{\princeton}{$^{17}$}
\newcommand{\ncsu}{$^{18}$}
\newcommand{\blhill}{$^{19}$}
\newcommand{\ttu}{$^{20}$}
\newcommand{\queens}{$^{21}$}
\newcommand{\tum}{$^{22}$}
\newcommand{\cs}{$^{,}$}
\newcommand{\clara}{$^{a}$}

\begin{document}

\title{Design improvements to cables and connectors in the \sc{Majorana Demonstrator}}

\author{C R~Haufe\unc\cs\tunl, A L~Reine\unc\cs\tunl, N~Abgrall\lbnl, S I ~Alvis\uw, I J~Arnquist\pnnl, F T~Avignone~III\usc\cs\ornl, A S~Barabash\itep, C J~Barton\usd, F E~Bertrand\ornl, T~Bode\mpi, A W~Bradley\lbnl, V~Brudanin\jinr, M~Busch\duke\cs\tunl, M~Buuck\uw, T S~Caldwell\unc\cs\tunl Y-D~Chan\lbnl, C D~Christofferson\sdsmt, P-H~Chu\lanl, C~Cuesta\uw\cs\clara, J A~Detwiler\uw, C~Dunagan\sdsmt, Yu~Efremenko\ut\cs\ornl, H~Ejiri\ou, S R~Elliott\lanl, T~Gilliss\unc\cs\tunl, G K~Giovanetti\princeton, M P~Green\ncsu\cs\tunl\cs\ornl, J~Gruszko\uw, I S~Guinn\uw, V E~Guiseppe\usc, L~Hehn\lbnl, R~Henning\unc\cs\tunl, E W~Hoppe\pnnl, M A~Howe\unc\cs\tunl, K J~Keeter\blhill, M F~Kidd\ttu, S I~Konovalov\itep, R T~Kouzes\pnnl, A M~Lopez\ut, R D~Martin\queens, R~Massarczyk\lanl, S J~Meijer\unc\cs\tunl, S~Mertens\mpi\cs\tum, J~Myslik\lbnl, C~O'Shaughnessy\unc\cs\tunl, G~Othman\unc\cs\tunl, W~Pettus\uw, A W P~Poon\lbnl, D C~Radford\ornl, J~Rager\unc\cs\tunl, K~Rielage\lanl, R G H~Robertson\uw, N W~Ruof\uw, B~Shanks\unc\cs\tunl, M~Shirchenko\jinr, A M~Suriano\sdsmt, D~Tedeschi\usc, J E~Trimble\unc\cs\tunl, R L~Varner\ornl, S~Vasilyev\jinr, K~Vetter\lbnl, K~Vorren\unc\cs\tunl, B R~White\lanl, J F~Wilkerson\unc\cs\tunl\cs\ornl, C~Wiseman\usc, W~Xu\usd, E~Yakushev\jinr, C-H~Yu\ornl, V~Yumatov\itep, I~Zhitnikov\jinr, and B X~Zhu\lanl}                                                          
\address{\unc Department of Physics and Astronomy, University of North Carolina, Chapel Hill, NC, USA}
\address{\tunl Triangle Universities Nuclear Laboratory, Durham, NC, USA}
\address{\lbnl Nuclear Science Division, Lawrence Berkeley National Laboratory, Berkeley, CA, USA}
\address{\uw Center for Experimental Nuclear Physics and Astrophysics, and Department of Physics, University of Washington, Seattle, WA, USA}
\address{\pnnl Pacific Northwest National Laboratory, Richland, WA, USA}
\address{\usc Department of Physics and Astronomy, University of South Carolina, Columbia, SC, USA}
\address{\ornl Oak Ridge National Laboratory, Oak Ridge, TN, USA}
\address{\itep National Research Center ``Kurchatov Institute'' Institute for Theoretical and Experimental Physics, Moscow, Russia}
\address{\usd Department of Physics, University of South Dakota, Vermillion, SD, USA} 
\address{\mpi Max-Planck-Institut f\"{u}r Physik, M\"{u}nchen, Germany}
\address{\jinr Joint Institute for Nuclear Research, Dubna, Russia}
\address{\duke Department of Physics, Duke University, Durham, NC, USA}
\address{\sdsmt South Dakota School of Mines and Technology, Rapid City, SD, USA}
\address{\lanl Los Alamos National Laboratory, Los Alamos, NM, USA}
\address{\ut Department of Physics and Astronomy, University of Tennessee, Knoxville, TN, USA}
\address{\ou Research Center for Nuclear Physics, Osaka University, Ibaraki, Osaka, Japan}
\address{\princeton Department of Physics, Princeton University, Princeton, NJ, USA}
\address{\ncsu Department of Physics, North Carolina State University, Raleigh, NC, USA}
\address{\blhill Department of Physics, Black Hills State University, Spearfish, SD, USA}
\address{\ttu Tennessee Tech University, Cookeville, TN, USA}
\address{\queens Department of Physics, Engineering Physics and Astronomy, Queen's University, Kingston, ON, Canada} 
\address{\tum Physik Department, Technische Universit\"{a}t, M\"{u}nchen, Germany}
\vspace{0.2cm}
\address{\clara Present Address: Centro de Investigaciones Energ\'{e}ticas, Medioambientales y Tecnol\'{o}gicas, CIEMAT, 28040, Madrid, Spain}
\vspace{0.2cm}

\ead{crhaufe@live.unc.edu | reineal@live.unc.edu}


\def\MJD{{\sc{Majorana Demonstrator}}}
\def\DEM{{\sc{Demonstrator}}}

\begin{abstract}
 The \MJD~is an experiment constructed to search for neutrinoless double- beta decays in germanium-76 and to demonstrate the feasibility to deploy a ton-scale experiment in a phased and modular fashion. It consists of two modular arrays of natural and 76Ge-enriched germanium p-type point contact detectors totaling 44.1 kg, located at the 4850’ level of the Sanford Underground Research Facility in Lead, South Dakota, USA.  The \DEM~uses custom high voltage cables to bias the detectors, as well as custom signal cables and connectors to read out the charge deposited at each detector’s point contact.  These low-mass cables and connectors must meet stringent radiopurity requirements while being subjected to thermal and mechanical stress. A number of issues have been identified with the currently installed cables and connectors. An improved set of cables and connectors for the \MJD~are being developed with the aim of increasing their overall reliability and connectivity. We will discuss some of the issues encountered with the current cables and connectors as well as our improved designs and their initial performance.
\end{abstract}

\section{Introduction}
The \MJD~is a neutrinoless double-beta decay experiment using Germanium as source and detector. The \DEM~contains 44.1-kg of Ge detectors divided between two independent cryostats \cite{general}. In total, the two modules contain 14.4 kg of \textsuperscript{nat}Ge and 29.7 kg of germanium enriched to 88\% Ge-76, the double beta decay isotope. The goals for the \DEM~are to demonstrate background levels low enough to justify building a tonne scale experiment, establish the feasibility of constructing and fielding modular arrays of Ge detectors, and search for additional physics beyond the Standard Model, such as solar axions and dark matter. The \DEM~is operating underground at the 4850’ level of the Sanford Underground Research Facility with the best energy resolution of any $0\nu\beta\beta$ experiment. Initial results based on datasets 3 and 4 indicate a 2.4 keV FWHM at 2039 keV and a projected background of $5.1^{+8.9}_{-3.2}$ c/(ROI-t-y), which is in good agreement with the \DEM's background goals \cite{tcald}.

Numerous measures are responsible for the \DEM's low background levels. In addition to the shielding provided by the rock overhead, the detector array is surrounded by a low-background passive Cu and Pb shield with an active muon veto. Ultra-low-activity components and construction techniques are also used to limit contaminants. In particular, the cryostats and other copper components were constructed using ultra-clean, electroformed copper. Current assay upper limits predict a background of $\leq 2.45$ counts/ROI-t-y based on the \DEM's achieved resolution\cite{general} \cite{assay}. In addition to these hardware-based background reduction techniques, the p-type point-contact detector design allows for optimal pulse shape discrimination to distinguish candidate double beta decay events from background events.  

The \DEM's background goal presents unique challenges in designing high voltage and signal cable systems. Cables and connectors must be kept as low mass as possible to limit radioactive backgrounds. Strict radiopurity requirements also control what materials can be used, meaning that standard commercial products are often not an option. Custom-made components were designed and implemented to meet these requirements, but connectivity problems and high voltage breakdowns have necessitated a redesign of some of these components. 

The \DEM~is currently operating 41 of 58 installed detectors.  7 of the non-operating detectors have problems associated with the signal connectors that are located on the cryostat cold plate or with damaged Low Mass Front End boards. The other 10 non-operating detectors cannot be electrically biased due to because of problems with HV cables, connectors, and in one instance a likely detector problem.  The improvements to cables and connectors discussed here are aimed at raising the percentage of operational detectors to $>90\%$.

\section{High Voltage Cables and Connectors}
In the \DEM, high voltage (HV) is applied to the outer contact of P-Point Contact (PPC) High Purity Germanium (HPGe) detectors.  An HV card supplies voltage to radiopure, in-vaccuum HV cables through custom pin connectors on a vacuum flange. Each HV cable carries this voltage to a detector through a custom electroformed Cu HV fork connected to a copper ring that makes contact with the outer surface of a detector, opposite the point contact.

The HV cable is constructed with a picocoax design, in which a central conductor is wrapped in a layer of FEP insulation, a tightly wound copper ground shield, and finally a second layer of FEP insulation that serves as the outer jacket.  These cables, manufactured by Axon', are rated to carry 5kV DC.  They exhibit a low linear mass density of 3 g/m and have an outer diameter of 1.2mm.  The Cu ground shield has a gauge of 50 AWG.

\begin{figure}[htb]
\begin{center}
\includegraphics[scale=0.35]{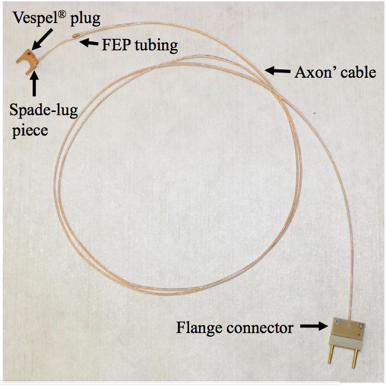}\\
\textbf{Figure 1:} \MJD~HV cable.  The copper HV fork is shown in the upper left-hand corner of the photo.  The flange connector is connected at the opposite end of the cross-arm at the vacuum flange.
\end{center}
\end{figure}
\smallskip
During initial operations multiple detectors exhibited HV ``breakdowns" in which there were significant discharges.  These detectors were fully or partially biased down to prevent damage to associated electronics.  It was determined the breakdowns were occuring between the central conductor and the outer ground shield.  These breakdowns were largely eliminated when the HV cable Cu ground shields were disconnected from ground. Of the detectors that are currently operating, 11 were brought on-line due to this change. 

A series of stress tests were performed on a sample HV cable to determine possible failure modes.  It was determined that kinked cables can lead to the same breakdown signatures observed in the \DEM~commissioning phase. The likely cause of HV breakdowns is a deformity in the layer of insulation separating the Cu ground shield from the central conductor due to kinked or crushed cables.  Damage to these cables likely occurred during installation, as no significant breakdowns were detected in cable testing following production and preceding installation.

The collaboration has encountered additional problems with the current design of the HV cables and connectors.  The Vespel clamp plug that covers the exposed end of the central conductor at the HV fork was found to not be secure for all detectors.  Additionally, collaborators have identified a risk of intermittent connection at the vacuum flange.

To address these issues, the collaboration plans on undertaking a full replacement of HV cables and connectors installed in the \DEM.  An existing set of Axon' HV cables will be installed with the same specifications as before.  To avoid the damaging of cables during installation, improved baffle plates will be set within the cross-arm to manage and direct cables to the detector cryostat.  Additionally, ePTFE thread will be used to bundle the cables together, providing further management and protection within the cross-arm.

Rather than using a Vespel clamp plug to cover the exposed end of the central conductor at the HV fork, a crimped connection will lock the central conductor in place with the HV fork, improving security.  A new set of PEEK connectors will be assembled to provide improved connectivity of the high voltage cable at vacuum flange, with new sockets that have a higher clamping force.

\section{Signal Cables and Connectors} 
The {\sc{Majorana}} signal cable and connector system is designed to transmit electronic pulses containing information about events in the germanium detectors. When an event occurs, charge is collected at the point contact and transmitted to a Low Mass Front End board (LMFE) with a FET that amplifies the signal\cite{guinn}. Each LMFE is connected to the preamp using four coaxial Axon' cables. Each set of four cables is divided into two separate cable bundles: one connecting the LMFE to a Vespel connector at the coldplate and another running between the coldplate and the D-sub connectors at the vacuum flange.

The Axon' signal cables have the same picocoax design as the HV cables described above. However, the signal cables have a smaller outer diameter of 0.4 mm, leading to a reduced linear mass density of 0.4 g/m. The cables have an impedance of 50 $\Omega$ and a capacitance of 87 pF/m. 

The main challenge presented by the {\sc{Majorana}} signal cable system is the difficulty of fabricating Vespel connectors that are robust enough to withstand temperature cycling without the use of conventional spring components that fail the \DEM's radiopurity requirements. The beryllium copper (BeCu) contacts used in many commercial connectors have unacceptably high \textsuperscript{232}Th and \textsuperscript{238}U activities. The Vespel connectors currently installed in the \DEM~are instead designed to avoid the need for contact springs, but this design requires very precise machining to ensure a secure connection. The machining constraints of the \DEM's underground machine shop have led to unreliable connectors.    

While the vacuum-side D-sub connectors are not responsible for observed connectivity issues in the \DEM, installation problems in the D-sub connectors reduced the number of viable spare channels. There is also evidence of damage to the signal cables during installation. Observed instances of electrical shorts between signal cable ground shields and the coldplate indicate problems with at least one detector's signal cables.  

In order to improve the reliability of Vespel connectors at the coldplate, the signal connector design has been modified to incorporate a fuzz button contact. These fuzz button contacts are manufactured out of gold-plated molybdenum wool by Custom Interconnects.  Unlike the BeCu contacts typically used to provide springiness in commercial connectors, fuzz buttons are expected to meet the \DEM's stringent radiopurity requirements based on assay results from the SuperCDMS collaboration \cite{CDMS}. The new connector design also provides a more secure connection using a mechanical locking mechanism. Prototypes of the improved connector design have passed 100\% of initial liquid nitrogen dunk tests, indicating that they will be able to withstand temperature cycling. A comparison of the old and new Vespel connector designs can be seen in Figure 2.  

The improved Vespel connectors will be installed during a replacement of the entire signal cable and connector system, planned to take place concurrently with the HV cable system upgrade. During this upgrade, the D-sub connectors at the vacuum flange will be replaced with more reliable commercial connectors from Glenair. Based on the evidence of damage to some existing signal cable ground shields, increased measures will be taken to  protect signal cables during installation. Like the HV cables, signal cables will be bundled using ePTFE thread. 

\begin{figure}[htb]
\begin{center}
\includegraphics[scale=0.6]{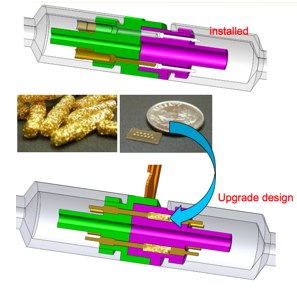}\\
\textbf{Figure 2:} Comparison of the installed and upgraded Vespel connector designs. A close-up of the fuzz button contacts and a size comparison are shown in the center images.  
\end{center}
\end{figure}

\section{Status and Outlook}
The upgrades to HV and signal cables and connectors discussed in sections 2 and 3 will undergo thorough testing.  A test stand using the \DEM~prototype cryostat will be used with a string of detectors to test upgraded HV cables and read out signal into upgraded signal cables.  An assay of the materials that will be used for the upgrade is also underway.  The manufactured cables to be used in the upgrade will be assembled with their corresponding connectors at UNC before shipment to the \DEM~site. Installation in Module 1 is scheduled to begin in the summer of 2018.

\section{Conclusion}
The \MJD~uses low-mass high voltage and signal cables that must meet stringent radiopurity, thermal stress, and mechanical stress requirement.  Issues with connectivity and stability have laid out an initiative for a cable and connector upgrade.

Thorough testing of all new high voltage and signal cables and connectors is underway. Upon completion of testing, cables assembled will be shipped to the \DEM~site at SURF. Data collection following the upgrade should commence sometime in Q4 2018. 

\ack
This material is based upon work supported by the U.S. Department of Energy, Office of Science, Office of Nuclear Physics, the Particle Astrophysics and Nuclear Physics Programs of the National Science Foundation, and the Sanford Underground Research Facility.   

\section*{References}

\smallskip

\end{document}